\documentclass[12pt,a4paper]{article}
\usepackage[utf8]{inputenc}
\usepackage[margin=1in]{geometry}
\usepackage{amsmath}
\usepackage{amsfonts}
\usepackage{amssymb}
\usepackage{graphicx}
\usepackage{cite}
\usepackage{url}
\usepackage{hyperref}
\usepackage{float}
\usepackage{booktabs}
\usepackage{longtable}

\title{Climate Risk Stress Testing in California: A Geospatial Framework for Banking and Climate-Exposed Sectors}
\author{
Satya Narayana Panda\\
Finance Department, University of New Haven\\
spand14@uunh.newhaven.edu
\and
Aishworzo Saha\\
Finance Department, University of New Haven
}
\date{April 2026}

\begin{document}

\maketitle

\begin{abstract}
This paper develops a geospatial framework for climate risk stress testing in California with applications to banking and climate-exposed sectors such as agriculture, real estate, and tourism. The study integrates physical hazard mapping, sector-specific exposure analysis, and scenario-based financial risk assessment to evaluate how wildfires, drought, flooding, extreme heat, and transition risks may affect regional economic activity and financial stability. The framework is intended to support portfolio monitoring, climate scenario analysis, and institutional readiness under emerging disclosure and risk-management standards. In addition, the paper provides a survey-based implementation guide for benchmarking current climate-risk practices and data needs across industry and academic stakeholders.

\end{abstract}

\section{Introduction}

Climate risk has moved from a long-horizon environmental concern to a core financial stability issue. For banks, investors, and public-finance stakeholders, the relevant question is no longer whether climate shocks matter, but how quickly those shocks can be transmitted into borrower performance, collateral values, local public finance, and asset repricing. California provides an especially informative setting for this question because it combines large and diversified capital markets with concentrated exposure to wildfire, drought, coastal flood risk, extreme heat, and regulatory transition dynamics \cite{wuebbles2017climate, california2018fourth}.

From a finance perspective, climate exposure is fundamentally spatial. Mortgage collateral, agricultural production, hospitality revenue, and municipal tax bases are all tied to location-specific hazards. A balance sheet that appears diversified at the sector level may still be highly concentrated when examined through geography or hazard type. Standard risk systems often capture borrower, sector, and macro factors, but do not always map those factors to granular physical exposure or to the interaction between physical and transition risks.

This paper develops a finance-centered framework for climate risk stress testing in California. The objective is not to present a finished reduced-form estimation exercise, but to articulate a research design and implementation architecture that is suitable for high-quality empirical work and institutional application. The paper makes three contributions. First, it organizes climate risk around the main channels that matter for financial institutions: cash-flow impairment, collateral repricing, operating disruption, and broader market spillovers. Second, it proposes a geospatial approach for linking county- and asset-level hazards to portfolio-level expected loss and valuation outcomes. Third, it outlines a practical empirical agenda and survey instrument for benchmarking current institutional preparedness across banking and adjacent climate-exposed sectors.

\section{Related Literature and Research Positioning}

The climate-finance literature has developed along three closely related lines. The first studies climate risk as a financial stability and disclosure problem. Policy and supervisory work emphasizes that physical and transition shocks can alter asset values, default probabilities, and market liquidity in ways that are not well captured by backward-looking risk models \cite{tcfd2017recommendations, battiston2017climate, cftc2020managing, federal2021principles}. The second line examines asset pricing and valuation effects, showing that climate exposure is increasingly capitalized into real estate values, discount rates, and expected returns \cite{bernstein2019climate, giglio2021climate, dietz2016climate}. The third line links climate shocks to regional economic performance, demonstrating that temperature, drought, and environmental catastrophes can depress productivity, incomes, and long-run growth \cite{kahn2019long, burke2015global, hsiang2017estimating}.

This paper sits at the intersection of those strands but takes a more explicitly place-based finance view. Rather than treating climate risk as a single macro shock, the framework begins with local hazard intensity and asks how that local shock propagates through loan performance, local cash flows, collateral valuation, and portfolio concentration. That perspective is particularly important for California, where wildfire exposure, water stress, and coastal vulnerability are highly uneven across counties and metropolitan areas \cite{westerling2006warming, abatzoglou2016impact, california2018fourth}. 

The paper also extends the discussion beyond banks alone. Agriculture, real estate, tourism, and local public finance are not peripheral sectors in the California economy; they are central transmission mechanisms through which climate shocks can affect lenders and investors. Framing the problem this way creates a more useful bridge between academic finance, prudential supervision, and applied institutional risk management.

\section{Framework and Methodological Architecture}

The proposed framework has three layers: \emph{hazard measurement}, \emph{financial exposure mapping}, and \emph{balance-sheet transmission}. Let $g$ denote a geographic unit such as a ZIP code or county, $i$ an exposure or instrument, and $s$ a climate scenario. The first layer measures physical hazards $H_{g,s}$ such as wildfire intensity, drought severity, flood probability, and heat stress. The second layer maps those hazards into the location and sectoral composition of loans, properties, businesses, and local fiscal bases. The third layer translates those exposures into expected loss, repricing, and capital strain.

For credit portfolios, the natural starting point is scenario-contingent expected loss:
\begin{align}
EL_{i,s} = PD_{i,s} \times LGD_{i,s} \times EAD_i,
\end{align}
where the probability of default and loss given default are permitted to vary with both physical and transition risk. A parsimonious empirical specification is:
\begin{align}
PD_{i,s} = PD_i^{0} \exp \left( \beta_1 H_{g,s} + \beta_2 T_{j,s} + \beta_3 U_g - \beta_4 A_i \right),
\end{align}
where $T_{j,s}$ captures transition risk in sector $j$, $U_g$ denotes local economic fragility, and $A_i$ measures borrower or asset adaptation capacity. The same logic applies to collateral values and local market liquidity, both of which can amplify effective credit loss even when the initial climate shock is localized.

For valuation-sensitive portfolios, we define an illustrative climate stress metric:
\begin{align}
Climate\text{-}VaR_s = \sum_{i=1}^{N} w_i \Delta V_{i,s} + \lambda \sum_{i=1}^{N} EL_{i,s},
\end{align}
where $\Delta V_{i,s}$ denotes mark-to-market repricing under scenario $s$ and $\lambda$ converts credit impairment into an equivalent capital or valuation burden. This formulation is intentionally general: it is suitable for banks and investors, while remaining close to the way finance practitioners already think about risk aggregation.

\subsection{Scenario Design}

The scenario design should combine acute shocks and medium-horizon structural change. In the California setting, the most relevant scenarios include severe wildfire seasons, prolonged drought, coastal flooding, urban heat stress, and policy-induced transition adjustment. These scenarios should not be interpreted as point forecasts. Instead, they serve as disciplined stress narratives that allow institutions to examine concentration, resilience, and nonlinear loss propagation \cite{ipcc2021climate}.

\begin{table}[H]
\centering
\caption{Illustrative climate scenarios for financial stress testing}
\small
\begin{tabular}{@{}p{3cm}p{5cm}p{5cm}@{}}
\toprule
Scenario & Representative shock & Primary financial channel \\
\midrule
Orderly transition & Gradual policy tightening & Repricing of carbon-sensitive cash flows \\
Disorderly transition & Abrupt regulation and market repricing & Spread widening and valuation adjustments \\
Physical risk shock & Wildfire, flood, or drought event cluster & Higher defaults, cash-flow stress, and collateral impairment \\
Compound scenario & Physical shock plus tighter financing conditions & Joint pressure on capital and liquidity \\
\bottomrule
\end{tabular}
\end{table}

\subsection{Data Strategy}

A credible empirical implementation would merge climate and financial datasets at a sufficiently granular geographic level. Relevant inputs include NOAA and state climate records, CAL FIRE and FEMA hazard layers, county assessor and property transaction files, bank call reports, HMDA data, agricultural production statistics, and municipal finance disclosures. The research objective is to move from descriptive climate exposure to institution-specific measures of expected loss, repricing sensitivity, and concentration risk.

\section{California Transmission Channels}

California is an appropriate laboratory for climate-finance analysis because different hazards map into different balance-sheet exposures. Wildland-urban interface areas matter for residential mortgage credit, property valuation, and municipal revenue stability. The Central Valley matters for agricultural lending, water-intensive production, and employment-linked local demand. Coastal counties matter for commercial real estate, tourism cash flows, and long-duration property valuation. Urban heat corridors matter for utility stress, labor productivity, public health spending, and operating income volatility.

The key point is not simply that these regions are exposed, but that their exposure affects the financial system through distinct transmission mechanisms.

\subsection{Banks}

For banks, the most important channels are credit deterioration, collateral impairment, and concentration risk. Wildfire or flood exposure can weaken household and small-business cash flow directly; it can also reduce the recoverable value of collateral when adaptation costs rise or local market liquidity weakens. Agricultural lenders are exposed to water availability, input-cost volatility, and yield uncertainty. Commercial real estate lenders are exposed to repricing when adaptation capex rises or expected occupancy falls in climate-sensitive areas.

\subsection{Asset Managers and Municipal Investors}

Asset managers and municipal investors face a different but equally important problem: climate risk changes the expected durability of local tax bases, infrastructure demands, and discount rates. Counties that require sustained adaptation spending or repeated disaster recovery may experience a gradual weakening of credit quality even in the absence of a single catastrophic event. This is precisely the kind of medium-horizon risk that traditional market measures often understate.

\section{Institutional Implementation and Testable Predictions}

A top-tier institutional climate stress test should be decision-relevant rather than merely descriptive. At minimum, it should answer five questions: where the portfolio is exposed, which hazards matter, which counterparties are most fragile, how collateral and local market conditions interact, and what capital or pricing adjustment follows under alternative scenarios.

Three empirical predictions follow naturally from the framework. First, borrowers located in high wildfire, drought, or flood-intensity areas should exhibit higher credit spreads and weaker performance after controlling for borrower quality and local economic conditions. Second, collateral discounts should be stronger in areas where local climate stress materially weakens property-market liquidity, even when the underlying hazard changes slowly. Third, institutions with geographically concentrated exposures should face disproportionately large stress losses relative to institutions that appear similar in aggregate sector composition.

These predictions are well suited to modern empirical finance methods, including panel designs with fixed effects, matched geographic comparisons, event studies around climate-policy or regulatory changes, and portfolio-level scenario simulation. The contribution of the present paper is to organize those empirical tests within a coherent climate-finance architecture rather than to rely on ad hoc hazard indicators.

\section{Practical Implications for Finance}

The practical message for financial institutions is straightforward. A serious climate-risk program should not begin with broad ESG language; it should begin with exposure mapping, underwriting sensitivity, and collateral resilience. Institutions with California exposure should track geographic concentrations, re-estimate borrower and collateral sensitivity to local hazards, and explicitly incorporate scenario-dependent local frictions into stress testing. For investors, the relevant issue is not only catastrophe risk but also the repricing of long-duration cash flows under changing expectations about adaptation cost and local economic resilience.

The role of regulators is equally important. Disclosure regimes such as California SB 253 and SB 261 are useful insofar as they improve comparability and discipline around measurement. But disclosure alone is not enough. What matters for financial stability is whether institutions can convert climate information into capital planning, pricing, concentration management, and governance decisions. That is where geospatial climate stress testing becomes especially valuable.

\section{Conclusion}

Climate risk in California should be understood as a finance problem with geographic structure. Physical hazards, transition dynamics, and local market responses do not affect all institutions equally; they affect them through portfolios, counterparties, and collateral located in specific places. A credible research and policy response therefore requires more than generic disclosure. It requires a framework that maps local hazard intensity into expected loss, repricing, and capital vulnerability.

This paper provides such a framework and positions it for empirical implementation. Its central argument is that climate stress testing in finance becomes substantially more informative when it is explicitly geospatial, sector-aware, and tied to core balance-sheet metrics. That is the level of analysis required if climate risk is to be treated with the seriousness already accorded to credit, liquidity, and market risk.

\section*{Acknowledgments}
We acknowledge the contributions of climate scientists, financial risk practitioners, and policy experts who have advanced our understanding of climate-finance interactions.

\bibliographystyle{unsrt}

\begin{thebibliography}{99}

\bibitem{wuebbles2017climate}
Wuebbles, D.J., Fahey, D.W., Hibbard, K.A., et al. (2017). \textit{Climate Science Special Report: Fourth National Climate Assessment, Volume I}. U.S. Global Change Research Program, Washington, DC.

\bibitem{tcfd2017recommendations}
Task Force on Climate-related Financial Disclosures. (2017). \textit{Recommendations of the Task Force on Climate-related Financial Disclosures}. Financial Stability Board.

\bibitem{battiston2017climate}
Battiston, S., Mandel, A., Monasterolo, I., Schütze, F., \& Visentin, G. (2017). A climate stress-test of the financial system. \textit{Nature Climate Change}, 7(4), 283-288.

\bibitem{dafermos2018climate}
Dafermos, Y., Nikolaidi, M., \& Galanis, G. (2018). Climate change, financial stability and monetary policy. \textit{Ecological Economics}, 152, 219-234.

\bibitem{kahn2019long}
Kahn, M.E., Mohaddes, K., Ng, R.N., Pesaran, M.H., Raissi, M., \& Yang, J.C. (2019). Long-term macroeconomic effects of climate change: A cross-country analysis. \textit{Energy Economics}, 104, 105624.

\bibitem{burke2015global}
Burke, M., Hsiang, S.M., \& Miguel, E. (2015). Global non-linear effect of temperature on economic production. \textit{Nature}, 527(7577), 235-239.

\bibitem{westerling2006warming}
Westerling, A.L., Hidalgo, H.G., Cayan, D.R., \& Swetnam, T.W. (2006). Warming and earlier spring increase western US forest wildfire activity. \textit{Science}, 313(5789), 940-943.

\bibitem{abatzoglou2016impact}
Abatzoglou, J.T., \& Williams, A.P. (2016). Impact of anthropogenic climate change on wildfire across western US forests. \textit{Proceedings of the National Academy of Sciences}, 113(42), 11770-11775.

\bibitem{lobell2008prioritizing}
Lobell, D.B., Burke, M.B., Tebaldi, C., Mastrandrea, M.D., Falcon, W.P., \& Naylor, R.L. (2008). Prioritizing climate change adaptation needs for food security in 2030. \textit{Science}, 319(5863), 607-610.

\bibitem{schlenker2009nonlinear}
Schlenker, W., \& Roberts, M.J. (2009). Nonlinear temperature effects indicate severe damages to US crop yields under climate change. \textit{Proceedings of the National Academy of Sciences}, 106(37), 15594-15598.

\bibitem{bernstein2019climate}
Bernstein, A., Gustafson, M.T., \& Lewis, R. (2019). Disaster on the horizon: The price effect of sea level rise. \textit{Journal of Financial Economics}, 134(2), 253-272.

\bibitem{giglio2021climate}
Giglio, S., Maggiori, M., Rao, K., Stroebel, J., \& Weber, A. (2021). Climate change and long-run discount rates: Evidence from real estate. \textit{The Review of Financial Studies}, 34(8), 3527-3571.

\bibitem{scott2012tourism}
Scott, D., Hall, C.M., \& Gössling, S. (2012). Tourism and climate change: Impacts, adaptation and mitigation. \textit{Routledge}.

\bibitem{gossling2012tourism}
Gössling, S., Scott, D., Hall, C.M., Ceron, J.P., \& Dubois, G. (2012). Consumer behaviour and demand response of tourists to climate change. \textit{Annals of Tourism Research}, 39(1), 36-58.

\bibitem{dietz2016climate}
Dietz, S., Bowen, A., Dixon, C., \& Gradwell, P. (2016). Climate value at risk of global financial assets. \textit{Nature Climate Change}, 6(7), 676-679.

\bibitem{ipcc2021climate}
IPCC. (2021). \textit{Climate Change 2021: The Physical Science Basis. Contribution of Working Group I to the Sixth Assessment Report of the Intergovernmental Panel on Climate Change}. Cambridge University Press.

\bibitem{cftc2020managing}
U.S. Commodity Futures Trading Commission. (2020). \textit{Managing Climate Risk in the U.S. Financial System}. Market Risk Advisory Committee, Climate-Related Market Risk Subcommittee.

\bibitem{federal2021principles}
Federal Reserve System. (2021). \textit{Principles for Climate-Related Financial Risk Management for Large Financial Institutions}. Board of Governors of the Federal Reserve System.

\bibitem{moody2021climate}
Moody's Analytics. (2021). \textit{Climate Change: The Physical Risks to Banks}. Moody's Corporation.

\bibitem{mckinsey2020climate}
McKinsey Global Institute. (2020). \textit{Climate risk and response: Physical hazards and socioeconomic impacts}. McKinsey \& Company.

\bibitem{hsiang2017estimating}
Hsiang, S., Kopp, R., Jina, A., Rising, J., Delgado, M., Mohan, S., ... \& Houser, T. (2017). Estimating economic damage from climate change in the United States. \textit{Science}, 356(6345), 1362-1369.

\bibitem{california2018fourth}
California Governor's Office of Planning and Research. (2018). \textit{California's Fourth Climate Change Assessment}. State of California.

\bibitem{yellen2021climate}
Yellen, J.L. (2021). Climate change poses serious emerging risks to the financial system. \textit{Brookings Papers on Economic Activity}, 2021(1), 1-78.

\end{thebibliography}

\appendix

\section{Supplementary Survey Instrument}

A full practitioner survey and implementation guide is provided in the companion supplementary document. The instrument is designed to collect evidence on five dimensions that are highly relevant for empirical finance research: (i) current climate-risk governance, (ii) the use of geospatial data in underwriting and portfolio surveillance, (iii) sectoral concentration and scenario design, (iv) operational and disclosure constraints, and (v) institutional readiness for climate stress testing.

\end{document}